\documentclass[%
 aip,
 jcp,
 amsmath,amssymb,
 reprint,%
]{revtex4-1}

\usepackage{amsmath, amsfonts}
\usepackage[version=3]{mhchem}
\usepackage[left=1.5cm, right=1.5cm, top=1.785cm, bottom=2.0cm]{geometry}
\usepackage{mathptmx}
\usepackage{commath}
\usepackage{braket}
\usepackage{graphicx}
\usepackage[format=plain,justification=justified,singlelinecheck=false,font={stretch=1.125,small,sf},labelfont=bf,labelsep=space]{caption}
\usepackage{float}
\usepackage[english]{babel}
\usepackage{array}
\usepackage[T1]{fontenc}
\usepackage[usenames,dvipsnames]{xcolor}
\usepackage{setspace}
\usepackage[colorlinks=true]{hyperref}

\usepackage{multirow}
\usepackage{bm}
\usepackage[version=3]{mhchem}

\begin{document}

\title{Increasing ion yield circular dichroism in femtosecond photoionisation
  using optimal control theory}

\author{Manel Mondelo-Martell}
\email{manel.mondelo@chemie.uni-frankfurt.de}
\altaffiliation[Current address:]
{Institut f\"ur Physikalische und Theoretische Chemie, J.W. Goethe Universit\"at
Frankfurt}
\affiliation{Dahlem Center of Complex Quantum Systems \& Department of Physics, Freie Universit\"at Berlin, Berlin.}
\author{Daniel Basilewitsch}
\email{daniel.basilewitsch@uibk.ac.at}
\affiliation{Dahlem Center of Complex Quantum Systems \& Department of Physics, Freie Universit\"at Berlin, Berlin.}
\altaffiliation[Current address: Institut f\"ur Theoretische
Physik, Universit\"at Innsbruck.]
{Institut f\"ur Physikalische und Theoretische Chemie, J.W. Goethe Universit\"at
Frankfurt}
\author{Hendrike Braun}
\email{braun@physik.uni-kassel.de}
\affiliation{Institute of Physics, Universit\"at Kassel, Kassel.}
\author{Christiane P. Koch}
\email{christiane.koch@fu-berlin.de}
\affiliation{Dahlem Center of Complex Quantum Systems \& Department of Physics, Freie Universit\"at Berlin, Berlin.}
\author{Daniel M. Reich}
\email{danreich@zedat.fu-berlin.de}
\affiliation{Dahlem Center of Complex Quantum Systems \& Department of Physics, Freie Universit\"at Berlin, Berlin.}

\begin{abstract}
We investigate how optimal control theory can be used to
improve Circular Dichroism (CD) signals for A--band of fenchone measured via the
photoionization yield upon further excitation. 
These transitions are electric dipole forbidden
to first order, which translates into low population transfer to the excited
state ($\approx 8\%$) but also allows for a clearer interplay between
electric and magnetic transition dipole moments, which are of the same order of
magnitude.
Using a model including the electronic ground and excited A state as well as all
permanent and transition multipole moments up to the
electric quadrupole, we find that the absolute CD signal of randomly oriented molecules
  can be increased by a factor \textcolor{black}{3.5} when using shaped laser
  pulses, \textcolor{black}{with the anisotropy parameter $g$ increasing from 0.06 to 1}.
Our insights provide additional evidence on how optimal control
can assist in amplifying chiral signatures \emph{via} interactions of permanent
and transition multipole moments.
\end{abstract}


\maketitle

\section{Introduction}
Chirality is the property of an object to not be superimposable 
with its mirror image through a combination of translations and rotations. 
In molecular systems, the mirrored forms, called enantiomers, have almost
entirely identical physical properties and interact indistinguishably
with non-chiral probes. 
At the same time, enantiomers can behave very differently in their interaction
with other chiral objects, as evidenced by the role of chirality in
many biochemical and medical processes. The development of better techniques for chiral
discrimination is therefore a very active field of research both from a theoretical
and experimental point of view.

Characterisation of chirality can be achieved \emph{via} chiral observables, \emph{i.e.} 
properties which take on different values
for each enantiomer. These techniques either rely on the interaction of the
sample with a \emph{chiral probe} or on the construction of a \emph{chiral setup}
to record the response\cite{Ordonez2018}.
A prototypical chiral observables is circular dichroism (CD), which
has been the subject of many theoretical \cite{Gunde1996,Jansik2005,
Stener2006,Ma2006,Rizzo2008,Horsch2011,Kroner2015} and experimental
\cite{Pulm1997,Boesl2006,Li2006,Bornschlegl2007,Breunig2009} studies. 
It is defined as the difference in absorption of circularly
polarised light (acting as the chiral probe) by the two enantiomeric forms of a chiral molecule.
To leading
order, circular dichroism is formed by the interplay between electric dipole and
magnetic dipole transitions.
Due to the generally low magnitude of magnetic dipole transition moments, CD
is a comparatively weak effect, amounting to less than 1\% of the total
absorption signal. In recent years a lot of effort has been invested in the
description and measurement of chiral observables which do not require 
involvement of the weak magnetic transition dipole moments, two notable 
examples being the photoelectron circular dichroism (PECD) and rotational spectroscopy 
with microwave three-wave mixing (M3WM). The enantiomeric contrast
obtained with these techniques reaches values of several percent even with 
transform-limited pulses.
Recently, it has been shown that optimal control theory can be used to increase
this value even further by exploiting interference between various photoionization
pathways\cite{Goetz2019}. 
For instance,
perfect anisotropy in the photoelectron angular distribution of a randomly 
oriented ensemble can be generated by exploiting interferences between 
single-photon pathways and a manifold of resonantly enhanced 
two-photon pathways \cite{Goetz2019a}.
Prospects for control are even more promising for M3WM where complete
enantiomer-specific population transfer is possible if a suitable combination of
frequencies and polarization for the electric fields driving the three-wave
mixing process is chosen\cite{Leibscher2020}.
These recent
successes in enhancing chiral signatures with shaped pulses
strongly suggest that interference between different excitation 
pathways may be a promising avenue to increase the contrast also in CD experiments.
However,
PECD and M3WM are pure electric dipole effects to first order, which leads to strong
transitions and the possibility to attain high contrast with moderate laser
intensities.
Conversely, CD relies on small magnetic dipole transition moments.
This raises the question whether interference effects between different excitation paths can
also be exploited to increase the contrast of the overall much weaker CD signals.
In the pursuit of understanding how to maximise the dichroic signal, the magnitude of
CD both as a function of laser pulse frequency \cite{Horsch2011,Kroner2015}, as
well as duration and envelope \cite{Ma2006} have been theoretically
investigated.  The majority of these studies focused on the leading-order
contribution to CD which involves only the electric dipole and magnetic dipole
transition moment in the absorption process.  However, it has long been
established that all multipolar terms in the light-matter interaction contribute
to CD \cite{Meath1987}. Indeed,
the electric quadrupole has a noticeable effect in the absorption signatures for
the 1,2-propylene oxide molecule when multiphoton excitations are considered\cite{Kroner2015}.
Even beyond chiral observables, there is recent interest in the study of
nondipole effect for many different physical processes, for example in
photoionisation \cite{Brennecke2018,Brennecke2018a, Hartung2021,Maurer2021} and
high-harmonic generation \cite{Gorlach2020,Jensen2021}. 

When attempting to increase the CD signal in an experiment, the final puzzle
piece 
is to transfer the optimal pulses from theory to the lab. This step requires to
disentangle
the physically relevant pulse properties from purely numerical
features that are often introduced by optimisation algorithms.
It also relies on an appropriate correspondence between the theoretical figure
of merit used in the optimization and the experimentally measured quantity.
Although experimental determination of absorption CD in the liquid
phase is well-established\cite{Berova2000}, optimal control of chiral
signatures for molecules in the gas phases presents a more adequate framework to
compare theory and experiment.
This is because emerging gas phase techniques allow for
measurements under collision- and interaction-free conditions
\cite{Zehnacker2010,Patterson2014} also in table-top setups, which 
therefore serve as the focal point of our investigations.
One way to assess CD is
mapping into the ionisation continuum: By using resonance enhanced mulitphoton
ionisation (REMPI), the helicity-dependent population of the optically active
electronic state is translated into ion yields \cite{Boesl2006,Li2006}. In combination with
time-of-flight laser mass spectrometry it is possible not only to measure the CD
of the parent ion but also of the fragment ions. Successful experiments have
recently been reported for several chiral molecules \cite{Horsch2011,Boesl2013,Hong2014} exploiting the
advent of advanced techniques such as the measurement of differential photoion
CD \cite{Fehre2021} or twin-peak setups for improved statistics adapted to
femtosecond laser pulses \cite{Ring2021}. For resonant processes, ion-yield CD
and absorption CD are closely connected with the normalised difference in ion
yields which for a specific resonance is equal to the normalised CD in
extinction \cite{Boesl2013}.

In this paper we investigate, for the first time, in how far
optimal control can exploit the interaction of the molecule with light via the
transition electric dipole,
magnetic dipole, and electric quadrupole as well as permanent electric dipole
and quadrupole moments to enhance circular dichroism. In order to avoid concealment
of the magnetic-dipole dependent CD signal by strong electric dipole
transitions, we focus on the A--band $n\rightarrow\pi^{\ast}$ transition in fenchone.
This transition is electric dipole-forbidden to first
order\cite{Pulm1997} which allows multipolar signatures to come to the forefront.
By using an effective two-level description together with a physically motivated 
parametrisation of the laser pulse, we are able to elucidate the role of different 
multipole orders in the optimised protocols.
Moreover, we examine how the optimised
pulses address different molecular orientations
when
maximising CD for an orientationally averaged ensemble. 
To stay close to experimental realisation, we also ensure that the pulse parameters
are feasible in state-of-the-art table-top experiments in the femtosecond regime.

This paper is organised as follows: Section 2 introduces our theoretical model
of fenchone and the molecule's interaction with a laser pulse as well as
our control functional and algorithm. Section 3 presents the results from our
optimisations with a particular focus on the role of the permanent electric
dipole and electric quadrupole transition moments for the control protocols.
Finally, Section 4 concludes and presents an outlook for future investigations.

\section{Theoretical Framework}

Setting the stage for an optimal control problem can be condensed to three main
questions: How do we represent the relevant physical states and model dynamics of the
molecule under study? How do we encode the physical control target in
a mathematical functional? And finally, which algorithm do we use to minimise,
respectively maximise, the target functional?  We begin by addressing the
question of representation and dynamics. To this end, in Section~\ref{ssec:light_chiral} we
introduce the description of the light-matter interaction of a laser field
with a chiral molecule beyond the electric dipole
approximation\cite{Krems2018,Milonni2019}. Then, we discuss the
most important features of the A--band transition in fenchone\cite{Pulm1997} in
Section~\ref{ssec:model}, which allows us to employ a minimal description for
the molecule that still contains all of the relevant physics. Specifically, we
motivate a model involving only two electronic states (the ground state and the
first excited state) and neglecting any additional degrees of freedom. Such
a two-level description does not account for continuum dynamics, but the
absorption step serves as an important first step towards optimising ion-yield
CD experiments -- a high contrast during the absorption step will lead to high
contrast in the ion yield.
Finally in Section~\ref{ssec:oct} we detail how to account for orientational averaging in
optimisations \cite{Goerz2014a},
introduce an optimisation functional
specifically adapted to the task of maximising CD, and discuss which algorithm is particularly
suitable for computing optimised pulses.

\subsection{Light-matter interaction in chiral molecules}\label{ssec:light_chiral}
Within the Born-Oppenheimer approximation, the Hamiltonian describing the
interaction of a molecule with an electromagnetic field using minimum coupling
is given by\cite{Meath1987, Bernadotte2012a, Krems2018, Milonni2019},
\begin{align}
  \begin{split}
  \hat{H}=&%
         -\sum_{j=1}^{N}
         \frac{1}{2m_e}\left(\hat{\bm{p}}_j-e\bm{A}(\hat{\bm{r}}_j,t)\right)^{2}
         -\frac{ge}{2m_e}\sum_{j=1}^{N}\bm{B}(\hat{\bm{r}}_j,t)\cdot \hat{s}_j\\
         & -\sum_i\sum_j \frac{Z_ie^2}%
         {4\pi\epsilon_0\left|\hat{\bm{R}}_i-\hat{\bm{r}}_j\right|}
        +\sum_i\sum_{j>i} \frac{e^2}%
         {4\pi\epsilon_0\left|\hat{\bm{r}}_i-\hat{\bm{r}}_j\right|}
  \end{split}\\
  \begin{split}
    =&\sum_{j=1}^{N}\frac{\hat{\bm{p}}_j}{2m_e}-\sum_{j=1}^{N} \frac{e}{m_e}\mathbf{A}(\hat{\bm{r}}_j,t)\cdot\hat{\bm{p}}_j\\
         & -\frac{e^2}{2m_e}\bm{A}^2(\hat{\bm{r}}_j,t)
          -\frac{ge}{2m_e}\sum_{j=1}^{N}\bm{B}(\hat{\bm{r}}_j,t)\cdot \hat{s}_j\\
         & -\sum_i\sum_j \frac{Z_ie^2}%
         {4\pi\epsilon_0\left|\hat{\bm{R}}_i-\hat{\bm{r}}_j\right|}
        +\sum_i\sum_{j>i} \frac{e^2}%
         {4\pi\epsilon_0\left|\hat{\bm{r}}_i-\hat{\bm{r}}_j\right|}.
  \end{split}
  \label{eq:full_molH}
\end{align}
\textcolor{black}{In Eq.~\eqref{eq:full_molH}, $\hat{\bm{p}}_j$, $\hat{\bm{r}}_j$ and $\hat{s}_j$
are the momentum, position and spin operators for the $j$th electron,
$\hat{\bm{R}}_i$ and $Z_i$ are the position operator and nuclear
charge for the $i$th nuclei,
$\bm{A}(\hat{\bm{r}}_j,t)$ is the vector potential, and
$\bm{B}(\hat{\bm{r}}_j,t)$ the magnetic field. Moreover, the constants $e$,
$m_e$, $g$ and $\epsilon_0$ correspond to the charge and mass of
the electron, the spin $g$-factor and the vacuum permittivity.}
The terms containing squares of the vector potential $\bm{A}$ can be safely
neglected outside the strong-field regime. More specifically,
for optical or near UV wavelengths, this approximation
is well-motivated for intensities $I<10^{18} \mathit{W}/\mathit{cm^2}$\cite{Ludwig2014}.
Introducing the expansion of the electric field (see Eqs~\eqref{eq:E_decomp}
and~\eqref{eq:Ex_appr} in Appendix~\ref{sec:light}), and performing a suitable
gauge transformation, the multipolar form of the light-matter
interaction Hamiltonian becomes\cite{Buckingham1959,Milonni2019} for an incident light field
propagating in z direction,
\begin{align}
  \begin{split}
    \hat{H}=\hat{H}_0&-|\varepsilon_x(t)|e^{i\varphi_x(t)}\hat{\mu}_x  -|\varepsilon_y(t)|e^{i\varphi_y(t)}\hat{\mu}_y\\
          & -\frac{\hat{Q}_{xz}}{c}\frac{d|\varepsilon_x(t)|e^{i\varphi_x(t)}}{dt}
          -\hat{m}_yB_y(t)\\
          & - \frac{\hat{Q}_{yz}}{c}\frac{d|\varepsilon_y(t)|e^{i\varphi_y(t)}}{dt}+\hat{m}_xB_x(t)\\
          &+\sum_{j=1}^{N}\bm{B}(\hat{\bm{r}}_j,t)\cdot \hat{s}_j,
  \end{split}
  \label{eq:multip_H}
\end{align}
where we collected the field-free terms into the time-independent Hamiltonian $\hat{H}_0$,
and used the definitions:
\begin{align}
  \hat{\mu}_{\alpha}&=\sum_{j=1}^{N}e\hat{\alpha}_j\\
  \hat{m}_{\beta}&=\sum_{j=1}^{N}\frac{e}{2m_e}\left(\hat{p}_{\alpha,j}\hat{\gamma}_j-\hat{\alpha}_j\hat{p}_{\gamma,j}\right)\\
  \color{black}\hat{Q}_{\alpha,\beta}&\color{black}=\sum_{j=1}^{N}\frac{e}{3}\hat{\alpha}_j\hat{\beta}_j-r^2\delta_{\alpha,\beta}
\end{align}
for electric dipole, magnetic dipole, and electric quadrupole operators
with $\alpha, \beta, \gamma \in \{x,y,z\}$
The first line in Eq.~\eqref{eq:multip_H} corresponds to the well-known dipole
approximation. It is equivalent to neglecting
the spatial dependence of the electric field entirely, such that only a function
of time remains.
Note that the dipole approximation removes any information 
concerning the direction of propagation,
$\bm{k}$, hence the handedness of circularly
polarised light is lost in such a model. As such, the only spatial
information encoded in the dipole approximation is the transition dipole moment $\bm{\mu}$ (a
molecular vector) and the plane of polarisation of light (a field pseudovector).
In order to get a chiral observable in the dipole approximation it is necessary
to introduce another vector in the process, so that we can define a pseudoscalar
that codifies the handedness of the molecule\cite{Ordonez2018}. 
For instance, the photoelectron angular
distribution of a randomly oriented sample of chiral molecules presents a
forward--backward asymmetry, known as \emph{Photoelectron Circular Dichroism}
(PECD)\cite{Ritchie1976}. The high contrast of the signal (up to 10\% between both enantiomers)
has motivated extensive theoretical and experimental studies.
Although PECD measurements provide comparatively high signal strengths, the
description of the corresponding observable is more complex than CD from a theoretical
point of view due to the necessity to describe the electronic continuum. 
Conversely, chiral signatures from light absorption - (conventional) CD
and ion-yield CD - primarily rely on bound-state
electronic properties.
Nevertheless, a chiral signature due to CD requires the helicity of 
light to explicitly enter the interaction \emph{via} the propagation
vector $\bm{k}$. For this reason our model includes the next-higher 
order term of the multipole expansion beyond the electric dipole, cf.~ the
second and third line of Eq.~\eqref{eq:multip_H}.

\subsection{System under study: A--band of fenchone}\label{ssec:model}
Electric dipole transitions are typically much stronger
than the corresponding magnetic dipole transitions. As a result, CD signatures
can easily be concealed by the electric dipole, leading to low-contrast signals in experiments. In order to avoid
such concealment, we focus
on the A--band of fenchone. This transition is electric dipole forbidden to
first order, since its main component is a symmetry
forbidden $n\rightarrow\pi^{\ast}$ transition\cite{Pulm1997}, and therefore
features
electric and magnetic transition dipole moments of the same order of
magnitude.

We seek to optimise laser pulses as used in table-top experiments, i.e. 
pulse lengths of the order of 100~fs and laser wavelengths
of $300$~nm. This timescale is significantly shorter than the
rotational periods of fenchone, which are of the order of 1~ns\cite{Loru2016}.
Therefore we can safely assume that the molecule
remains at a single fixed orientation during the full length of the pulse.
Conversely, the main vibrational modes of the fenchone molecule have periods of the order of 
50~fs. These short amplitude motions, however, correspond to
individual \ce{C-H} and \ce{C-C} bonds in the molecular backbone, and are not
expected to play a significant role in the electronic dynamics\cite{Longhi2006}. Therefore, we
will restrict the modeling to the electronic degree of freedom.

Representing the Hamiltonian, Eq.~\eqref{eq:multip_H}, in the
basis of electronic eigenstates of fenchone, $\ket{\Psi_n}=\ket{n}$, we
obtain the expression
\begin{align}
  \begin{split}
    \Braket{m|\hat{H}|n}=\Braket{m|\hat{H}_0|n}&\textcolor{black}{-}|\varepsilon_x(t)|e^{-i\varphi_x(t)}
                   \left(\Braket{m|\hat{\mu}_x|n}+\frac{1}{c}\Braket{m|\hat{m}_y|n}\right)\\
         &\textcolor{black}{-}|\varepsilon_y(t)|e^{-i\varphi_y(t)}\left(\Braket{m|\hat{\mu}_y|n}-\frac{1}{c}\Braket{m|\hat{m}_x|n}\right)\\
         &\textcolor{black}{-}\frac{1}{c}\frac{\dif|\varepsilon_x(t)|}{\dif t}e^{i\varphi_x(t)}\Braket{m|\hat{Q}_{xz}|n}\\
         &-\frac{1}{c}\frac{\dif|\varepsilon_y(t)|}{\dif t}e^{i\varphi_y(t)}\Braket{m|\hat{Q}_{yz}|n}\\
         \label{eq:Hmatrix}
  \end{split}
\end{align}
where we have used the fact that, for symmetry reasons, any contribution
due to spin vanishes for real-valued wave functions of singlet
states\cite{Bernadotte2012a,Krems2018}. \textcolor{black}{In Eq.~\eqref{eq:Hmatrix},
$|\varepsilon_{\alpha}(t)|e^{-i\varphi_{\alpha}(t)}$ is the $\alpha$ component of the
complex-valued Fourier transform of the electric field (see Appendix~\ref{sec:light} 
for details on the expansion of the electric field).}

We have calculated electronic state energies as well as permanent and transition moments
with the \textsc{Dalton 2020} software package
\cite{Aidas2014,DALTON2020} at Coupled Cluster Singles Doubles
(CCSD) level with a 6-31G basis set, employing the Linear Response theory
implementations described in Refs.~\citenum{Christiansen1996,Christiansen1998a,
Halkier1998}.
Due to the localised nature of the two states involved in the A--band transition (the ground and the
first electronic excited state), a more extended basis set
describing the strong Rydberg nature of higher excited states\cite{Goetz2017},
was not necessary.
Furthermore, in order to guarantee a good representation of the two states
in our minimal model, we included up to the fifth electronic excited
state when calculating electronic energies and multipole moments.
All computed quantities relevant for the optimisations
are provided in Table~\ref{tab:dalton}. Note that the permanent magnetic dipole moment is
  neglected due to the singlet nature of the electronic states considered.
\begin{table*}
  \centering
  \caption{Energies, permanent electric dipole and transition
  multipole moments for the ground and first electronic excited state of
  fenchone obtained at CCSD/6-31G level with
  \textsc{Dalton2020.0}.}
  \label{tab:dalton}
  \begin{tabular}{c|cccc|cccc}
    &\multicolumn{4}{c}{$\Ket{0}$}&\multicolumn{4}{c}{$\Ket{1}$}\\\hline
    &Energy & El. dip.& Mag. dip. & El. quad.& Energy & El. dip.& Mag. dip. & El. quad.\\
    &$\mathit{eV}$ &$\mathit{e a_0}$& $\mathit{e\hbar m_e^{-1}}$& $\mathit{e a_0^{2}}$&$\mathit{eV}$ &$\mathit{e a_0}$& $\mathit{e\hbar m_e^{-1}}$& $\mathit{e a_0^{2}}$\\\hline
    $\Bra{0}$& 0 &
                   {$\!\begin{aligned}
                    \begin{pmatrix}
                      -0.047 \\
                      -1.061 \\
                      -0.414
                    \end{pmatrix}
                    \end{aligned}$}
                    &
                    &
                   {$\!\begin{aligned}
                    \begin{pmatrix}
                      4.159 & 0.012 & -0.241\\
                      -0.012 & -5.841 & -3.411 \\
                      -0.240 & -3.411 & 1.682
                    \end{pmatrix}
                    \end{aligned}$}
                    &
                    &
                   {$\!\begin{aligned}
                    \begin{pmatrix}
                      0.0033 \\
                      0.0002 \\
                      0.0037
                    \end{pmatrix}
                    \end{aligned}$}
                    &
                   {$\!\begin{aligned}
                    \begin{pmatrix}
                      0.0851 \\
                      0.958 \\
                      0.426
                    \end{pmatrix}
                    \end{aligned}$}
                    &
                   {$\!\begin{aligned}
                    \begin{pmatrix}
                      0.003 & 0.110 & 0.221\\
                      0.110 & 0.026 & 0.052 \\
                      0.221 & 0.052 & 0.029
                    \end{pmatrix}
                    \end{aligned}$}
                    \\
    $\Bra{1}$&     &
                   {$\!\begin{aligned}
                    \begin{pmatrix}
                      0.003 \\
                      0.0002 \\
                      0.004
                    \end{pmatrix}
                    \end{aligned}$}
                    &
                   {$\!\begin{aligned}
                    \begin{pmatrix}
                      0.085 \\
                      0.958 \\
                      0.426
                    \end{pmatrix}
                    \end{aligned}$}
                    &
                   {$\!\begin{aligned}
                    \begin{pmatrix}
                      0.003 & 0.110 & 0.22\\
                      0.110 & 0.026 & 0.052 \\
                      0.221 & 0.052 & 0.029
                    \end{pmatrix}
                    \end{aligned}$}
                    &
                    4.01
                    &
                   {$\!\begin{aligned}
                    \begin{pmatrix}
                      -0.076 \\
                      -0.823 \\
                      -0.343
                    \end{pmatrix}
                    \end{aligned}$}
                    &
                    &
                   {$\!\begin{aligned}
                    \begin{pmatrix}
                      4.060 &-0.561 & 0.157\\
                      -0.561 & -4.512 & -1.943 \\
                      0.157 & -1.943 & 0.453
                    \end{pmatrix}
                    \end{aligned}$}
                    \\
  \end{tabular}
\end{table*}

\subsection{Optimal Control of Circular Dichroism}\label{ssec:oct}

Since we neglect rotational motion, our Hamiltonian
(Eqs.~\eqref{eq:multip_H} and \eqref{eq:Hmatrix})
only describes a single orientation of the chiral molecule with respect
to the light pulse. However, experiments are typically carried out
with a statistical ensemble of randomly oriented molecules which have to be
accounted for in our model.
\begin{figure}
  \centering
  \includegraphics[width=\columnwidth]{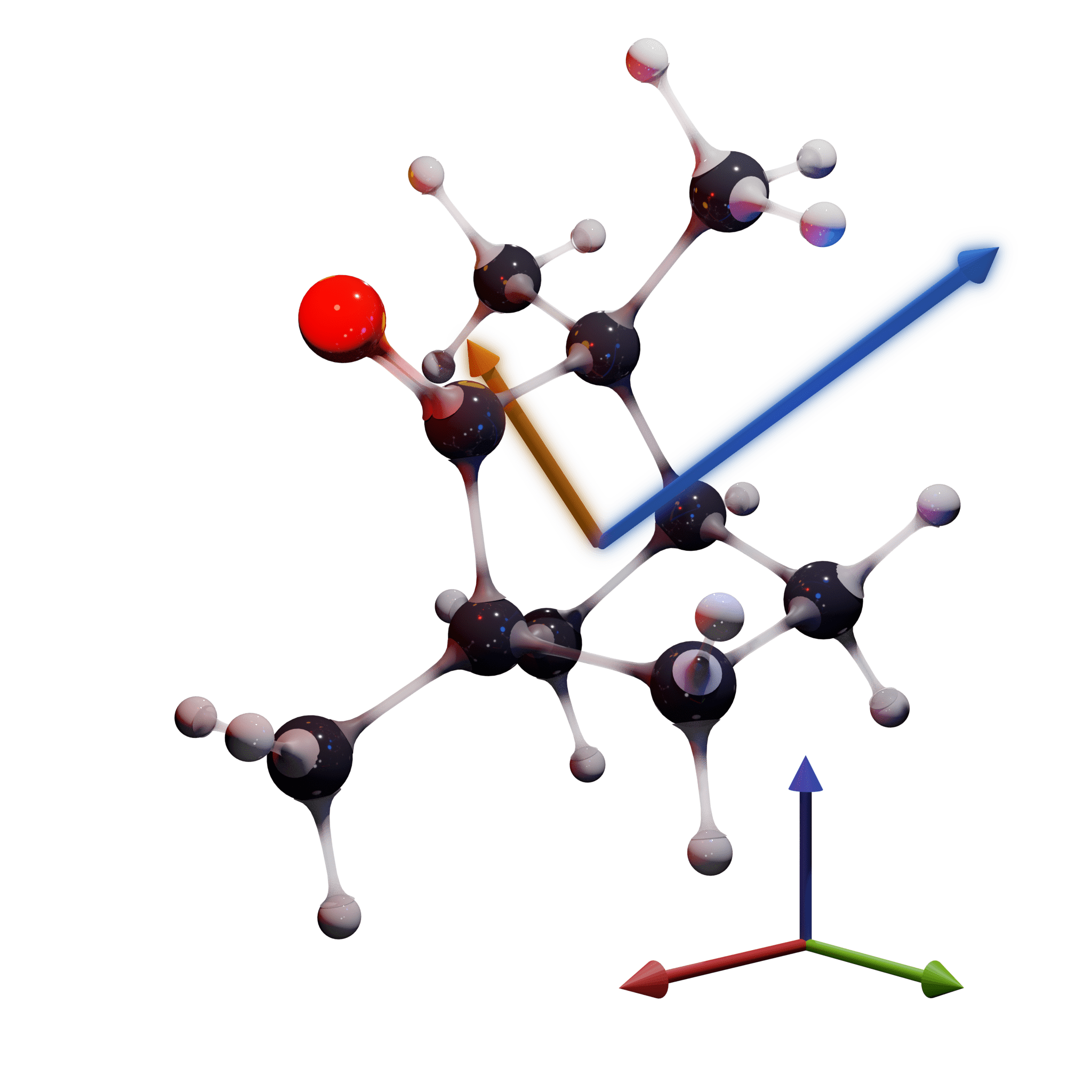}
  \caption{Reference geometry of fenchone, as obtained after optimisation at
  CCSD/6-31G level with \textsc{Dalton2020.0}, superimposed with the transition electric dipole
  moment (scaled $\times400$, orange) and transition magnetic dipole (scaled
  $\times 4$, blue). The coordinate system indicates the orientation of the
  molecular frame, with the RGB axes corresponding to the $x$, $y$, $z$ Cartesian
  coordinates}\label{fig:fenchone}
\end{figure}
Averaging over all Euler angles, defined with respect to the orientation shown
in Figure~\ref{fig:fenchone} in the $y-z-y$ convention, we obtain
for the excited state population of a single enantiomer
\begin{align}
  \begin{split}
    \left|\Braket{\Psi^1|\Psi_R(T)}\right|^{2}=\frac{1}{8\pi^2}\int_{0}^{2\pi}\int_{0}^{\pi}\int_{0}^{2\pi}
     &\left|\Braket{\Psi^1(\alpha,\beta,\gamma)|\Psi_R(\alpha,\beta,\gamma,T)}\right|^2\\
     &\sin{\beta}\dif\alpha\dif\beta\dif\gamma,
     \label{eq:rotaver}
  \end{split}
\end{align}
with $\ket{\Psi^{1}(\alpha,\beta,\gamma)}$ the excited state electronic eigenfunction, and
$\ket{\Psi_R(\alpha,\beta\gamma,T)}$ the
state of the $R$ enantiomer wave function at time $T$.

With this final puzzle piece on the question of representation in place,
we can now turn to address the description of the system dynamics. To this end, we employ the 
time--dependent Schr\"odinger equation,
\begin{align}
  \frac{\dif \Ket{\Psi(\bm{x},t})}{\dif
  t}=\frac{1}{i\hbar}\hat{H}(t)\Ket{\Psi(\bm{x},t_0)}.\label{eq:TDSE}
\end{align}
Although this equation of motion only describes coherent dynamics, such
a treatment is justified by the fact that any decoherence or decay 
is expected to occur on much longer time scales than the $fs$
pulse durations.
As it is commonly done in the field of optimal control, we have separated the 
Hamiltonian in Eq.~\eqref{eq:multip_H} into a field-free, time-independent \emph{system} 
Hamiltonian, $\hat{H}_0$, (the so-called \emph{drift})
and the time-dependent Hamiltonian due to the interaction
of the chiral molecule with the light field (the so-called \emph{control}
\textcolor{black}{$\varepsilon$}),
\begin{align}
  \hat{H}(t)=\hat{H}_0+\sum_{k=1}^N \varepsilon_k(t)\hat{H}_k.
\end{align}
The coupling to an external field provides a means to steer the dynamics of the system
towards a specific target, in our case by shaping the incident laser pulse.

Next, we address the question of how to encode the physical target in terms of a functional, $J_T$.
This functional quantifies how well the control pulses implement the optimisation goal.
In our case we target a high contrast in electronic state populations of
the two enantiomers of fenchone. We also expect this to serve as a precursor 
of getting high-contrast in ion-yield CD signals, since these experiments
only rely on the measurement of the number of ions obtained after absorption and ionisation.
As seen in the previous section, the two enantiomeric forms share the
same drift Hamiltonian, however, they feature a different relation between the
electric and magnetic dipole transition moments: for one enantiomer the product
of both quantities is positive, while for the other it is negative. This
relative arrangement of the multipole moments in the light-matter interaction term
is the source of the different dynamics in a given enantiomer when exposed to
a light source with different helicity, and thus the origin of circular
dichroism.
Note that in this work we use a complementary (and equivalent) point of view
to evaluate the CD: instead of changing the helicity of the pulse, we consider
how a specific light field interacts with each of the enantiomers.
Therefore, starting from chiral molecules in the ground state, we seek a pulse which
selectively excite one of the enantiomers while leaving the opposite form
in the ground state.
Once a difference in electronic state population between the
two enantiomers is established, a second pulse can be used to selectively ionise from
the higher energy level, thus obtaining an increased ionisation CD signal.
For a single orientation this goal can be encoded by a so-called
state-to-state optimisation \emph{via}
the following functional,
\begin{align}
  J_T=1-\frac{1}{2}\left(\left|\Braket{\Psi^1|\Psi_R(T)}\right|^2+\left|\Braket{\Psi^0|\Psi_S(T)}\right|^2\right)\label{eq:JT}.
\end{align}
Similar functionals aiming to increase the distinguishability of two systems
with a single control are also prominent, e.g., in
quantum discrimination of magnetic fields \cite{Ansel2018, vanDamme2018,Basilewitsch2020}.
In Eq.~\eqref{eq:JT}, $\Psi^0$ and $\Psi^1$ refer respectively to the ground
and electronic excited state of the chiral molecule. $\Psi_R(T)$
($\Psi_S(T)$) denotes the state of the $R$ ($S$) enantiomer at final time $T$.
This functional takes on its minimal value 0 when the
R enantiomer is completely excited and the S enantiomer remains entirely in the
ground state, and its maximal value 1 in the opposite scenario. Note that both extrema
correspond to perfect distinguishability, while a vanishing chiral
signal corresponds to a functional value of 0.5. Thus, increasing the distance
to this middle point, which can be achieved by either minimisation or maximisation,
improves the realisation of our physical goal.

The linearly polarised components of the electric field, $E_x(t)$
and $E_y(t)$, can be represented as two different control pulses which are optimised
independently allowing for arbitrary elliptical polarisation. Moreover, due to
the small absorption amplitude in the A--band transition of fenchone,
perturbative considerations are suitable to predict
which processes will be the most relevant in our simulations\cite{Meath1987}.
This suggests to employ a parametrisation for the control pulses, which allows
to reduce the dimensionality of the optimisation landscape.
Specifically, we expect three main contributions for each control field:
\textcolor{black}{DC components ($E_{x/y}^{(0)}$),
which primarily couple to the permanent dipole, one-photon components
($E_{x/y}^{(1)}$) with frequency $\omega^{(1)}$, which primarily couple to the electric
and magnetic dipole transitions, and two-photon components ($E_{x/y}^{(2)}$)
with frequency $\omega^{(2)}$ which primarily couple to the electric quadrupole
moments\cite{Meath1987}.
The interference between these couplings leads
to different excitation pathways, which will be the main resource exploited 
by the optimised pulses.
Following this physical intuition we parametrise our control field as
a superposition of the three aforementioned contributions:
\begin{align}
  E_x&=s(t)\left(E_x^{(0)}+E_x^{(1)}\sin{(\omega^{(1)} t)}+E_x^{(2)}\sin{(\omega^{(2)} t)}\right)\label{eq:pulse_params1}\\
  E_y&=s(t)\left(E_y^{(0)}+E_y^{(1)}\sin{(\omega^{(1)} t+\varphi)}+E_y^{(2)}\sin{(\omega^{(2)} t+\varphi)}\right),
  \label{eq:pulse_params2}
\end{align}
with $\varphi$ the relative phase between the $x$ and $y$ components of the electric
field.
In Eqs.~\eqref{eq:pulse_params1} and~\eqref{eq:pulse_params2} $s(t)$ is an
envelope function ensuring that the pulse is smoothly turned on and off. Here we
choose a squared sine as a good approximation to experimental pulse 
shapes\cite{Barth2009},
\begin{align}
  s(t)=\sin^2\left(\frac{\pi t}{T}\right).
\end{align}
}

Note that we keep $\omega^{(1)}$ as an optimisation parameter and do not fix it
to the resonant frequency of the electronic excitation ($\omega_r=4.01$~eV).
This is done to allow for non-resonant processes to be considered as candidates
for the optimal solutions and permits flexibility in view of potential DC and AC
Stark shifts. Conversely, we keep the frequency for the two-photon pathway to the
excited state fixed at $\omega^{(2)}=\omega^{(1)}/2$ to explicitly target
a bichromatic control mechanism using interference between a one- and two-photon
pathway. 

To account for orientational averaging
we perform an ensemble
optimisation\cite{Goerz2014a}: We propagate a set of differently orientated
molecules, each described by its own Hamiltonian, under the effect of the same
control pulses, and minimise the \emph{averaged} functional,
\begin{align}
  \begin{split}
    J_T^{\mathit{aver}}=\frac{1}{8\pi^2}\sum_{i=1}^{N_{\alpha}}\sum_{j=1}^{N_{\beta}}\sum_{k=1}^{N_{\gamma}}&
    \left[1-\frac{1}{2}\left(\left|\Braket{\Psi^1(\alpha,\beta\gamma,T)|\Psi_R(\alpha,\beta\gamma,T)}\right|^2\right.\right.\\
    &\left.\left.+\left|\Braket{\Psi^0(\alpha,\beta\gamma,T)|\Psi_S(\alpha,\beta,\gamma,T)}\right|^2\right)\right]\\
    &\sin{\beta}\Delta\alpha\Delta\beta\Delta\gamma.
  \end{split}
  \label{eq:JT_aver}
\end{align}
Note that we have replaced the integrals from Eq.~\eqref{eq:rotaver}
by sums, to account for the numerical necessity of discretising the set of orientations.
We have chosen $N_{\alpha}=N_{\gamma}=2N_{\beta}=14$
to sample the orientations equidistantly and with identical spacing for all three Euler angles, i.e.,
$\Delta=\Delta\alpha=\Delta\beta=\Delta\gamma$.

Since we can describe our control field with very few parameters,
cf.~Eqs~\eqref{eq:pulse_params1} and \eqref{eq:pulse_params2}, gradient-free optimisation methods are
particularly suitable. We have used \textcolor{black}{a combination of the Multi--Level Single
Linkage (MLSL) approach~\cite{RinnooyKan1987, RinnooyKan1987a, Kucherenko2005} and } the generalised simplex (or Nelder-Mead)
\cite{Nelder1965,Box1965,Richardson1973} algorithm as implemented in the python
\emph{NLopt} library\cite{Johnson}. \textcolor{black}{The MLSL algorithm
stochastically samples the parameter space of the optimisation. This global scan of the
optimisation landscape complements the local nature of the generalised simplex method and allows
to find the optimal solution even when several local minima are present. All
propagations have been performed using the \emph{QDYN} library\cite{qdyn}. }
Optimal control algorithms \emph{per se} do
not impose any restriction on the calculated pulses. However, in gradient free
methods it is easily possible to restrict the domain of the parameters to be optimised.
These constraints should be chosen
in order to obtain control pulses apt for experimental
applications, which are usually limited by the total pulse duration, and
the maximum field strength that can be generated.
In table-top setups, pulses with 30-40~fs duration and
peak electric field strengths of the order of $GV/m$ can be routinely obtained.
However, due to the small transition moments of the A--band
of fenchone, such pulses result in populations of the
excited state of around 1\%. These low values are insufficient to increase
the contrast in the CD signal with a high signal to noise ratio.
Preliminary simulations showed that we can obtain population transfer of
$\approx 10\%$ by using pulses of 100~fs and peak electric field
strength of 25.7~$\frac{\textrm{GV}}{\textrm{m}}$ (corresponding to a value of 
0.05 atomic units).
These restrictions are at the upper limit of experimental feasibility in
table top setups but are still possible, albeit challenging, to implement.
\textcolor{black}{Due to the dipole-forbidden nature of the A--band, peak
intensities of the order of $10^{14}\ W/cm^2$ result in a comparatively weak light-matter
interaction, as it can be seen from the fact that they only produce $\approx
10\%$ population transfer to the excited state. Therefore the pulses can still be considered
not to be in the strong field regime and we expect the assumptions of our two-state model to hold.
} It should be noted that a more complex model beyond a
two-level description would increase the dimension of the parameter space, where
gradient-based methods show their strengths. A more detailed discussion on which
optimisation algorithm is most suitable to a particular problem can be found for
example in Ref.~\citenum{Goerz2019}.

\section{Results and discussion}\label{sec:results}
\subsection{Circular dichroism of randomly oriented ensembles}\label{ssec:CDens}
\begin{figure*}
  \centering
  \includegraphics{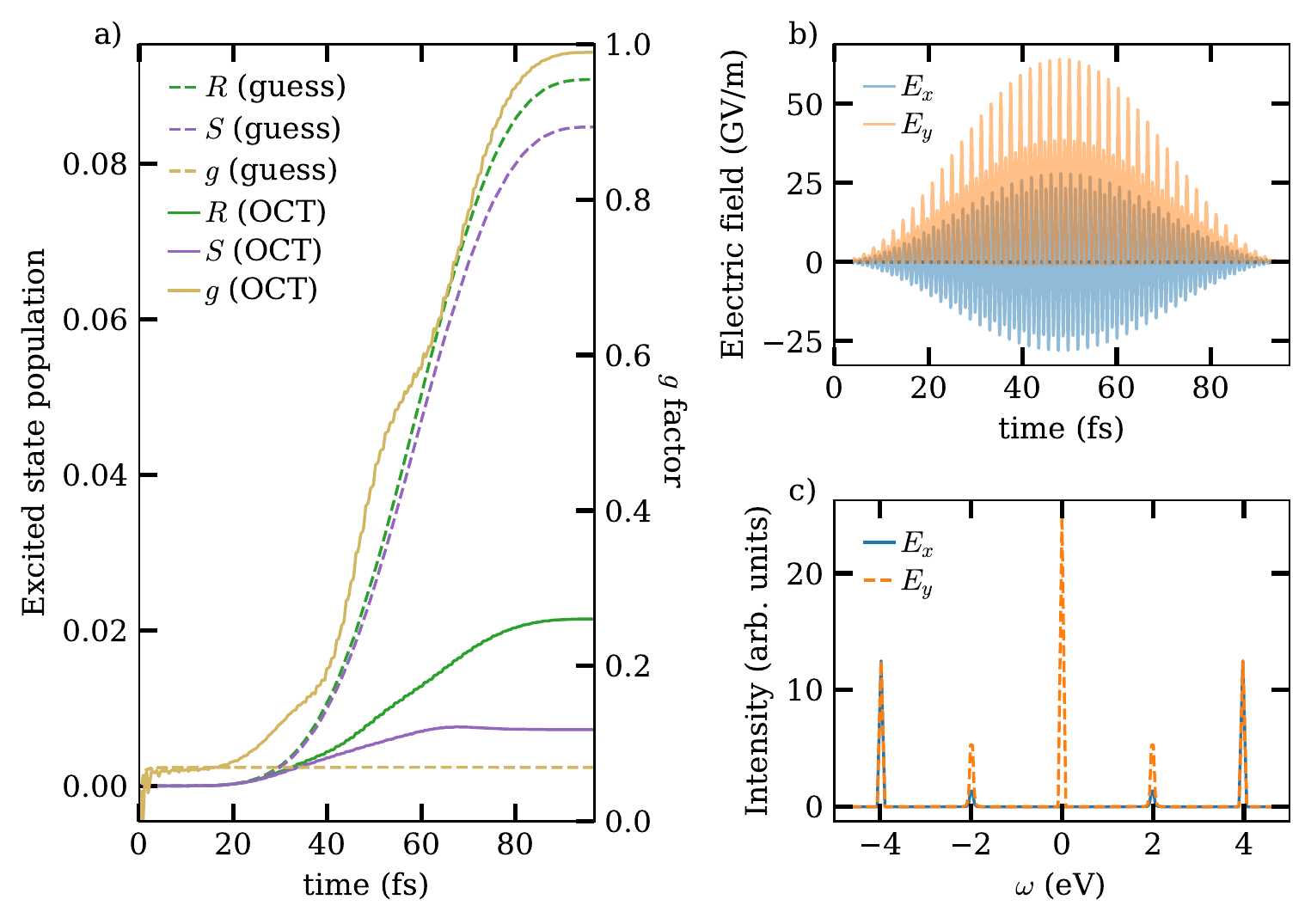}
  \caption{Results for the optimisation of circular dichroism of a rotational
  ensemble for the A--band
  transition of fenchone. a)
  Evolution of the excited state population as a function of time for the
  $R$ (green) and $S$ (purple) enantiomer of fenchone\textcolor{black}{, as well
  as the corresponding value of the anisotropy parameter $g$ (yellow, in the right $y$ axis)}. The dashed line
  corresponds to the circularly polarised guess pulse, while the solid line
  corresponds to the optimised control
  fields. \textcolor{black}{The oscillations of $g$ at short times are a numerical artifact due
  to the near 0 absorption of the excited states during the first
  femtoseconds.} b) Optimised pulses in time domain. c) Optimised pulses in frequency
  domain.}\label{fig:CDopt_orientaver_NM100fs}
\end{figure*}

As a guess for the optimisation we
choose a  50~fs FWHM circularly polarised pulse
with a single-frequency component at $\omega=4.01~\textrm{eV}$ and
$E_x^{(1)}=E_y^{(1)}=25.7\frac{\textrm{GV}}{\textrm{m}}$.
Due to the electric dipole forbidden nature of the
transition, population transfer to the excited state with this pulse reaches
maximum values of 8\%, with a difference in excited state population
between the R and S enantiomers 
around 0.5\%. We can quantify the dichroic signal with the anisotropy factor
$g$, defined as
\textcolor{black}{the ratio between the difference in absorption of circularly
polarised light between the left and right enantiomers over the absorption of
non-polarised light for that band, taken as the average absorption of both
enantiomers\cite{Kuhn1930}:
\begin{align}
  g=\frac{I_{\mathit{left}}-I_{\mathit{right}}}{\frac{1}{2}\left(I_{\mathit{left}}+I_{\mathit{right}}\right)},
\end{align}
}
where $I_\mathit{left}$ and $I_\mathit{right}$ refer to the absorption of
a given enantiomer\textcolor{black}{, which in our system can be taken as
equivalent to the excited state populations.}
Even though this definition is usually used for monochromatic, circularly
polarised pulses we will also employ it for our optimised field.
In the case of our guess circularly polarised pulse we obtain $g=6.25\cdot 10^{-2}$,
which compares very well with the $5\cdot 10^{-2}$ value reported in the literature\cite{Pulm1997}.
The generalised simplex
(or Nelder-Mead) algorithm minimises the value of the rotationally
averaged functional $J_T^{\mathit{aver}}$ (\emph{c.f.} Eq~\eqref{eq:JT_aver})
by independently varying the different components of the pulse:
the intensity of the DC component ($E_{x/y}^{(0)}$, coupling primarily to the
permanent electric dipole moment), the one-photon component
($E_{x/y}^{(1)}$, coupling primarily to
the electric and magnetic dipole transitions), the two-photon component
($E_{x/y}^{(2)}$, which coupling primarily to
electric quadrupole moment), as well as the frequency $\omega^{(1)}$.
The optimised pulses in time and frequency domain as well as the resulting population
dynamics are displayed in Figure~\ref{fig:CDopt_orientaver_NM100fs}. The
optimised values for the pulse parameters, cf.
Eq.~\eqref{eq:pulse_params1} and~\eqref{eq:pulse_params2}, are shown in Table~\ref{tab:NM100fs_params}.
\textcolor{black}{
\begin{table*}
  \centering
  \caption{Parameters of the circularly polarised guess pulse (anisotropy $g=6.25\cdot 10^{-2}$ after orientational averaging) and
    the optimised pulse (anisotropy $g=1.0$ after orientational averaging).}
  \label{tab:NM100fs_params}
  \begin{tabular}{c|ccc|cc|c|ccc|cc|c}
    &\multicolumn{6}{c}{Optimised pulse}&\multicolumn{6}{c}{Guess pulse}\\\hline
    &$E^{(0)}$&$E^{(1)}$&$E^{(2)}$&$\omega^{(1)}$ &$\omega^{(2)}$& $\varphi$
    &$E^{(0)}$&$E^{(1)}$&$E^{(2)}$&$\omega^{(1)}$ &$\omega^{(2)}$& $\varphi$\\
    &\multicolumn{3}{c|}{$\frac{GV}{m}$}&\multicolumn{2}{c|}{$\mathit{eV}$}&
    &\multicolumn{3}{c|}{$\frac{GV}{m}$}&\multicolumn{2}{c|}{$\mathit{eV}$}
    &\\\hline
    $E_x$ & 4.95$\cdot 10^{-3}$ & 27.71&3.26 &3.97 &1.99 &$\pi/2$ & 0.0 & 25.70& 0.0 & 4.01
    & - & $\pi/2$\\
    $E_y$  &25.71& 25.71 & 12.86 &3.97 &1.99 & & 0.0   & 25.70 & 0.0   & 4.01 & - & \\
  \end{tabular}
\end{table*}
}
\textcolor{black}{
Remarkably, the anisotropy obtained with the guess circularly polarised pulse reaches its
maximum value in the first few femtoseconds, and remains constant throughout the
rest of the dynamics. This stands in sharp contrast to the behaviour under the
optimised pulse, which shows a gradual increase of the anisotropy throughout the
whole pulse.
The parameters in Table~\ref{tab:NM100fs_params} also show that the optimisation
slightly alters the frequency compared to the resonance frequency $\omega_r$
from the guess pulse. We attribute this feature to the combined effects of AC
and DC Stark shifts induced by the field}. Moreover, the optimised pulse not
only addresses dipolar transitions at the frequency $\omega_r$ but also DC field
contributions ($\omega = 0$) due to the permanent electric dipole, and
two-photon \textcolor{black}{($\omega = \omega^{(1)}/2$)} contributions arising
from coupling to the electric quadrupole \textcolor{black}{(see also
Figure~\ref{fig:CDopt_orientaver_NM100fs}c). Interestingly, only the $y$ field
contributes to the DC component. This can be attributed to the symmetry of the
rotationally averaged system, which causes the maximum anisotropy to arise when
the DC component is aligned with only one polarisation vector.}
\textcolor{black}{
The resulting difference in excited-state population between the two
enantiomers, and thus the chiral contrast, increases by a factor 2.5 compared to
the guess field.  Moreover, the overall signal strength in the A--band is also
reduced from $\approx 9\%$ for the circularly polarised pulse to $\approx 2\%$
for the optimised one.  Given that the field strength of both sets of pulses is
similar, we can attribute the difference in dynamics to interference effects
between the different excitation pathways.  All in all, the combination of
increase of chiral contrast and decrease of overall absorption results in
a increase of the anisotropy parameter to almost $g=1.0$, i.e., absorption in
one enantiomer is almost doubled compared to its mirror image when using the
optimised pulse.}

\begin{figure*}
  \centering
  \includegraphics[width=\textwidth]{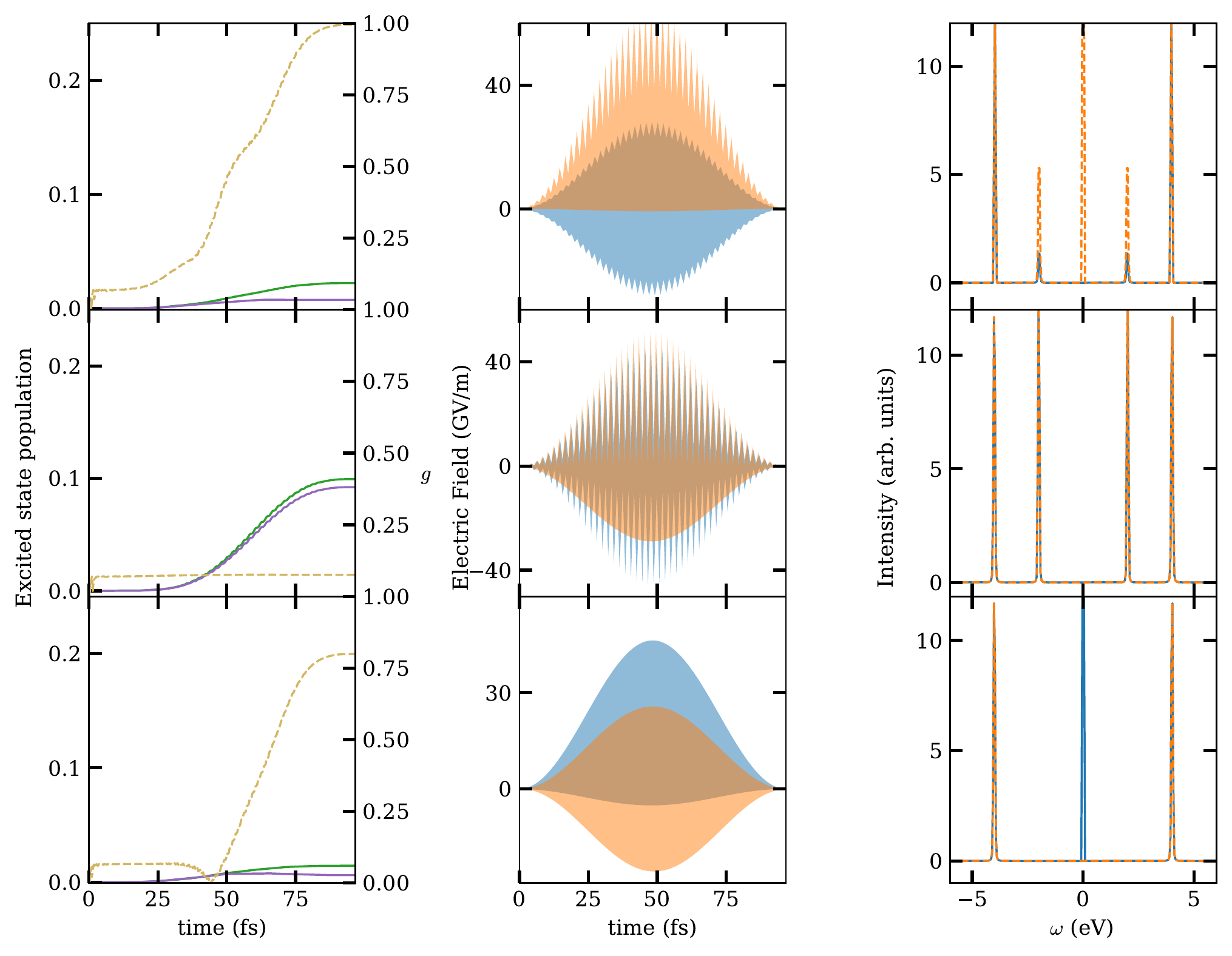}
  \caption{\textcolor{black}{Comparison of optimisations with a fully
  parametrised pulse (top), a restricted pulse without
  DC contribution (middle) and a restricted pulse without $\omega^{(2)}$ contribution (bottom).
  Left to right: excited
  state population for the R (solid green) and S (solid purple) enantiomer and anisotropy factor $g$ (dashed yellow, in right $y$ axis); envelope of the optimised pulses
  in time domain; optimised pulses in frequency domain ($E_x$ blue, $E_y$ orange).
  }}\label{fig:restr_comparison}
\end{figure*}

Despite the fact that the leading order for circular dichroism is usually given
by electric and magnetic dipole transitions, our optimisation results for the
(to first order) dipole-forbidden transition in fenchone reveals the significance 
of multipolar terms beyond the electric and magnetic dipole transition moment.
To further illustrate and investigate their significance we employ
the following two approaches: On the one hand, we use a \emph{restricted
model} by removing the corresponding coupling operator from the Hamiltonian and
perform another optimisation of the pulse parameters. On the other hand,
we optimised a \emph{restricted pulse} by simulating dynamics using the full
Hamiltonian, yet constraining 
$E_0=0$ (respectively $E_2=0$) for the control fields.
The comparison of the simulations with these different schemes is shown in
Figure~\ref{fig:restr_comparison}.
\textcolor{black}{From these different setups we can clearly see that
the permanent electric dipole, through its interaction with the
DC component of the electric field, is the critical ingredient for the 
success of the ensemble optimisation: 
An optimisation restricting the pulse to a vanishing zero-frequency component
does not significantly increase the anisotropy
with respect to the guess. Conversely, an optimisation using a restricted pulse
without two-photon contributions still offers a significant increase of the anisotropy
factor, but only reaches around 0.75 anistropy in contrast to the almost 1.00
when utilising all possible pathways.
Evidently, all multipolar terms in our model provide a
significant optimisation resource, since they open multiphoton excitation pathways
which allow to exploit interference effects towards the desired objective.
}

\subsection{Oriented Circular Dichroism}\label{ssec:OCD}
For typical experiments on circular dichroism in the gas phase the chiral signal is
orientationally averaged over all possible orientations of the molecular target.
From a theoretical point of view it is nevertheless interesting to also consider
how control pulses can induce different absorption between two enantiomers
for a single, space-fixed orientation.
Specifically, we first investigate optimisations for single orientations of
fenchone with respect to the light field. Then, we analysed how the optimal controls
obtained from ensemble optimisation act on individiual orientations. The
comparison of these two sets of simulations helps to gain
insight into the underlying control mechanism.

\begin{figure*}
  \includegraphics[width=\textwidth]{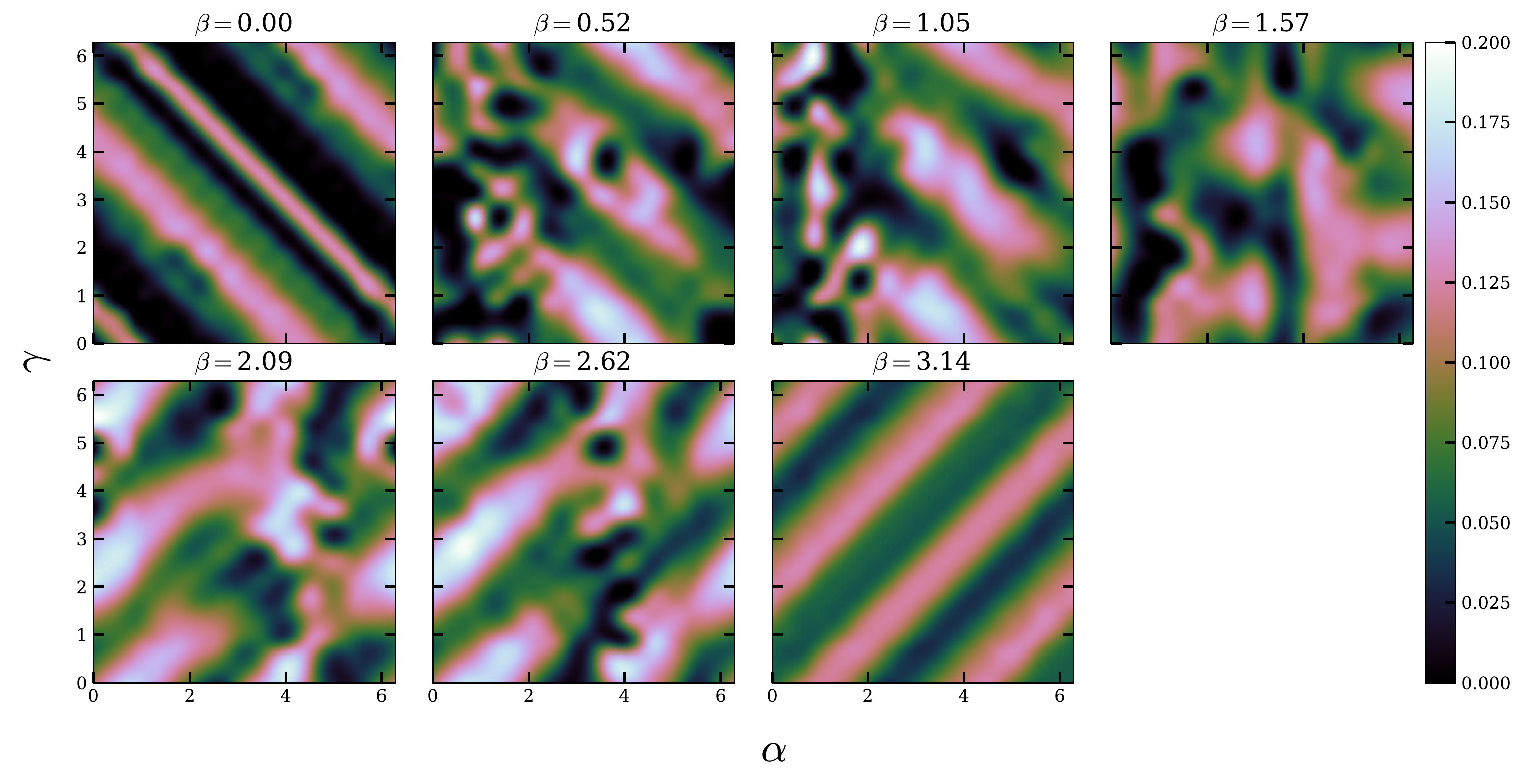}
  \caption{Value of the fidelity (Eq.~\eqref{eq:F}) optimised for individual orientations
  of fenchone as a function of the Euler angles $\alpha$, $\beta$ and $\gamma$.
  Lighter areas correspond to higher fidelities, \emph{i.e.} better
  chiral distinguishability.}\label{fig:EulerSurvey_indopt}
\end{figure*}

We analyse the population
dynamics for the two enantiomeric forms for a given orientation.
For one of the enantiomers the optimised pulse aims to minimise excited state
population transfer altogether, or at least to return all intermittent
population in the excited state back to the ground state at the end of the
pulse. At the same time, for the mirror image, the optimised field tries to maximise
population transfer to the excited state. The latter process (maximisation of population
in the excited state for one enantiomer) is limited by the available fluence
in the pulse. Specifically, taking into account our restrictions
on field strength and the limited pulse length (cf. Sec.~\ref{ssec:oct}), 
the optimisation does not have enough resources to get a complete population transfer to the
excited state.
Remarkably, not all orientations are equally easy to control
in terms of distinguishability. Every pulse-molecule geometry yields
different values for the components of the permanent dipole and
transition multipole moments in the control Hamiltonian, and small values of
these moments can prevent the optimisation of CD altogether.
Figure~\ref{fig:EulerSurvey_indopt} shows the fidelity $F$ of the optimisation
for different individual orientations of the system. Using the functional
$J_T$ defined in Eq.~\eqref{eq:JT}, this quantity is defined
as
\begin{align}
  F=2|0.5-J_T|\label{eq:F}
\end{align}
and takes the value 1 for perfectly distinguishable systems, and 0 for
completely indistinguishable ones, cf. our discussion in Sec.~\ref{ssec:oct}
A closer look at the values of the transition moments for different orientations
shows that the possibility for improvement via optimal control depends strongly on the
interplay of the different components of the vectors: For instance, in the
orientation $\alpha=1.35$, $\beta=2.62$, $\gamma=5.38$, where anisotrpy
appears not to be improvable, the value of the
electric dipole transition moment $x$ component is one order of magnitude
smaller than in the neighboring optimisable orientation $\alpha=0.45$,
$\beta=2.62$, $\gamma=5.38$. Similar relations, with one or more relevant
transition moments becoming small, can be observed for several other areas
that only show negligible improvement though optimisation.

\begin{figure*}
  \includegraphics[width=\textwidth]{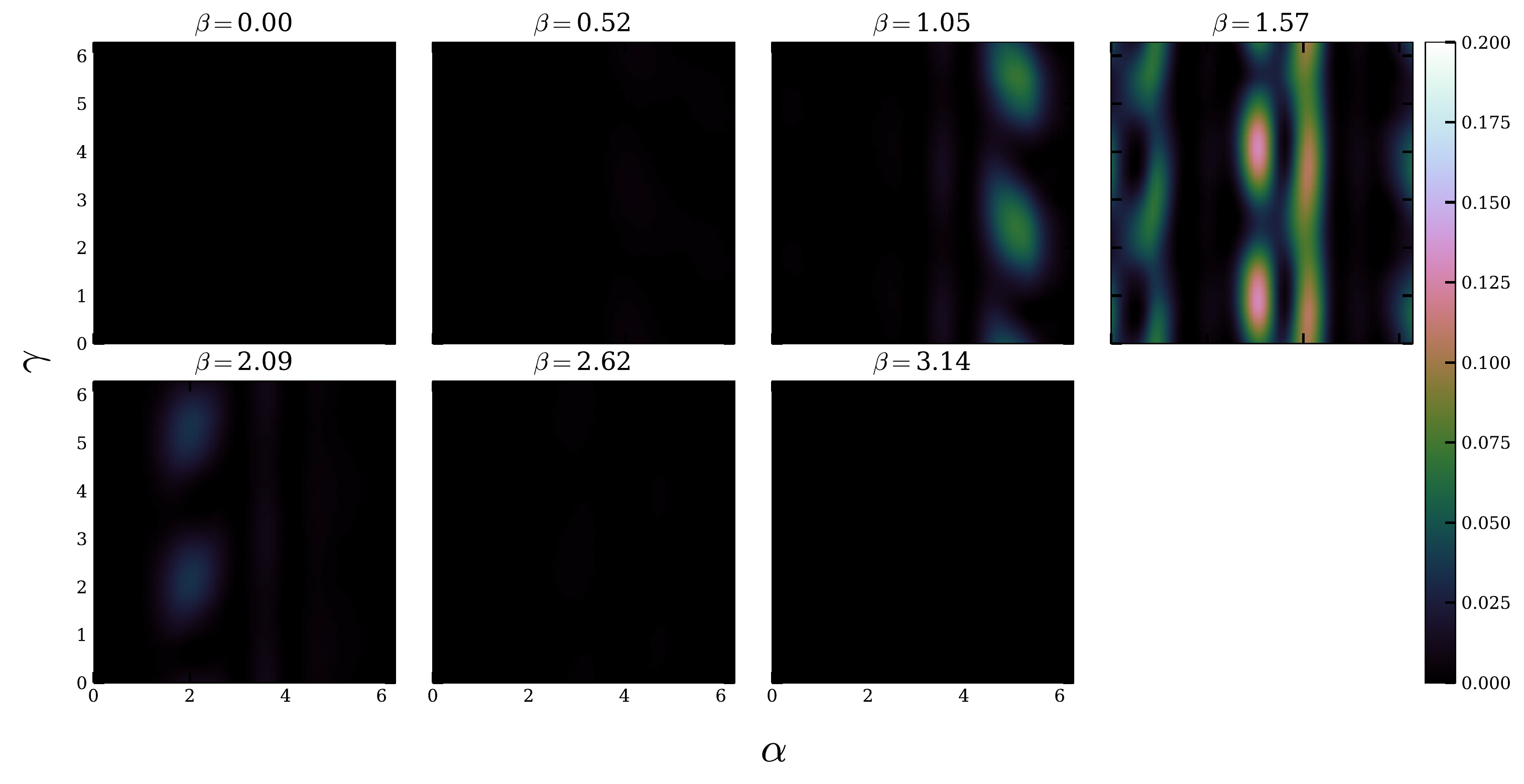}
  \caption{Value of the fidelity (Eq.~\eqref{eq:F} after irradiation with the ensemble optimised
  pulse (Table~\ref{tab:NM100fs_params} and
  Figure~\ref{fig:CDopt_orientaver_NM100fs}) as a function of the Euler angles
  $\alpha$, $\beta$ and $\gamma$. 
  Lighter areas correspond to higher fidelities, \emph{i.e.} better
  chiral distinguishability.}\label{fig:EulerSurvey_OrAverPulse}
\end{figure*}

The fidelity obtained by considering the action of the pulse optimised for a rotational average
(Figure~\ref{fig:CDopt_orientaver_NM100fs} and Eq~\eqref{eq:JT_aver}) on individual
orientations is shown in Figure~\ref{fig:EulerSurvey_OrAverPulse}.
\textcolor{black}{We can clearly see that 
the ensemble optimised pulse
increases the distinguishability for a subset of orientations in the region $1.5<\beta<2.7$
while having close to no effect on the rest. The reason for this is that due to
the weighting factor $\sin\beta$ appearing from the rotational
averaging (\emph{cf.} Eq.~\eqref{eq:rotaver}), orientations in that region
have an above average contribution to the ensemble.
This incentivises the optimisation algorithm to focus on this domain.
By comparison to Figure~\ref{fig:EulerSurvey_indopt} we can see that
the optimisation also targets those orientations
intrinsically more favourable in terms of distinguishability.}

\section{Summary and Conclusions}
We have shown that optimal control can be used to
increase the absorption contrast in the A--band of the two
enantiomers of fenchone by independently shaping the $x$ and $y$ components of
the incident light field.  In order to do so, we have developed a minimal
molecular model, including only the electronic ground and first excited state of
the molecule. 
Our model consistently includes all light-matter interaction terms up to one order
beyond the dipole approximation, \emph{i.e.} the electric and magnetic dipole
transition moments (which are the leading-order contribution to CD), electric
quadrupole moment, and permanent electric dipole moment. 
The magnetic and electric dipole moments, including the permanent electric
dipole, contribute appreciably to the excitation dynamics, while the
electric quadrupole only has a minor effect.
We have obtained optimised pulses that increase the orientationally averaged
contrast in the excited state population between the two enantiomers by
\textcolor{black}{almost a factor of twenty compared to
a monochromatic circularly polarised pulse, while also decreasing the overall
absorption to around a quarter compared to the guess pulse. These effects are
a result of the interferences between the different excitation paths generated
by the optimised pulses, which feature spectral contributions with frequency
$\omega^{(1)}$ and $\omega^{(1)}/2$
with $\omega^{(1)}\approx \omega_r$, 
as well as a DC field component for the electric field. The DC component proves to
be critical for the optimisation, while the $\omega_r/2$ contribution, 
coupling primarily to the quadrupole, has a smaller yet still clearly noticeable effect.}
As a result, we have shown that it is possible to achieve
control for CD signatures by exploiting different multipolar contributions
of the light-matter interaction, even in a basic two-level description.
While such a description simplifies the electronic structure to only the ground 
and a single excited state, our model still captures most of the
relevant dynamics for table-top pulses in the femtosecond regime.

To rationalise the results of the ensemble optimisation, we have studied how the
optimised pulse affects specific orientations of the fenchone molecule. We have observed
that only a subset of geometries shows an increase in the population difference
between ground and excited state compared to the guess pulse.
In order to explain this behaviour, we have performed full optimisations on
individual orientations sampling the whole rotational space. The optimisation results
show that the regions where the rotational ensemble optimised pulse performs
better correspond to domains in which the optimisation of individual orientations
is more favourable. This is related to a stronger coupling, and hence an enhanced
addressability, by virtue of larger overlaps between the molecular
transition moments and the electromagnetic field.

In a next step, this knowledge is to be transferred to the experiment. Instrumental
restrictions will influence the implementation of the optimised pulses:
Our pulse lengths and peak intensities are, albeit challenging, attainable in
state-of-the-art table-top setups, but our optimised solutions also prominently
feature a DC component for the electric field which may be problematic for
an experimental implementation.
Our optimisations show, that attempting to 
increase the CD signal with a more restricted protocol (\emph{i.e.} removing the
DC component of the field) leads to only marginal increase of the
distinguishability, pointing towards the critical role the DC field plays for
our optimisations. 
Several further avenues towards obtaining more easily 
realisable yet efficient pulses can be considered. A first option is to add more electronic levels in
our model. This would add more excitation pathways that can be addressed
simultaneously by a multicolored laser pulse. The interference between these
pathways is expected to lead to better control mechanisms similarly to the case
of PECD\cite{Goetz2019, Goetz2019a}. 
Secondly, we have observed that the excited state population difference can be easily
increased for particular orientations of the molecule with respect to the light
pulse.
This suggests that a pre-pulse which
induces a partial orientation of the molecular ensemble might be a promising
strategy\cite{Koch2019}. Moreover, it is conceivable to engineer an optimised pulse
which both orients and excites the chiral molecules. Although such a study would
require a description of different timescales to account for rotational dynamics,
recent advances in controlling the rotational state of chiral molecules show
a lot of promise in that direction\cite{Tutunnikov2021}.

\section*{Conflicts of interest}
There are no conflicts to declare.

\section*{Acknowledgements}
We would like to thank Thomas Baumert and Andr\'es Ordo\~nez
for helpful discussions, and Marec Heger for providing the molecular model from
Figure~\ref{fig:fenchone}. Financial support by the Deutsche
Forschungsgemeinschaft (DFG, German Research Foundation)—Projektnummer
328961117—SFB ELCH 1319 is gratefully acknowledged.

\bibliography{main.bib}

\appendix
\section{Light-Matter interaction beyond the Electric Dipole Approximation}
\label{sec:light}
Circular dichroism is formed by the interplay between different electric and
magnetic interaction terms, to first order the electric dipole and the magnetic
dipole. Magnetic dipole transitions are usually much weaker than electric dipole transitions
and comparable in strength to electric quadrupole transitions. To adequately 
describe all relevant orders in CD, we move beyond the commonly used electric dipole 
approximation and take into account the next-highest orders in the form of magnetic dipole 
effects and electric quadrupole effects.
To this end, the electric field of an electromagnetic wave propagating
along the $z$ axis is written as a superposition of plane waves,
\begin{align}
    \bm{E}(\bm{r},t) =
    \frac{1}{\sqrt{2\pi}}\int_{0}^{\infty} \dif \omega
                       \left(\bm{\varepsilon}^*(\omega)e^{-i\frac{\omega}{c}\hat{\bm{e}}_z\cdot\bm{r}-\omega t}
                        +\bm{\varepsilon}(\omega)e^{i\frac{\omega}{c}\hat{\bm{e}}_z\cdot\bm{r}-\omega t}\right).
                        \label{eq:Evec_gen}
\end{align}
In this expression, the Fourier coefficients $\bm{\varepsilon}(\omega)$ are vector
quantities which describe the polarisation in the $xy$ plane.
By projecting the electric field onto the cartesian unit vectors
$\bm{\hat{e}}_x$ and $\bm{\hat{e}}_y$, Eq.~\eqref{eq:Evec_gen} can be brought
into a more familiar form,
\begin{align}
  \bm{E}(\bm{r},t) = E_x(\bm{r},t)\hat{\bm{e}}_x+
                             E_y(\bm{r},t)\hat{\bm{e}}_y
                             \label{eq:E_decomp}
\end{align}
with
\begin{align}
  \begin{split}
  E_x(\bm{r},t) =
  \frac{1}{\sqrt{2\pi}}\int_{0}^{\infty} \dif \omega
                     &\left(|\varepsilon_x(\omega)|e^{-i\varphi_x}\right.
                     e^{-i\frac{\omega}{c}\hat{\bm{e}}_z\cdot\bm{r}}
                     e^{i\omega t}\\
                     & +|\varepsilon_x(\omega)|e^{i\varphi_x}
                      e^{i\frac{\omega}{c}\hat{\bm{e}}_z\cdot\bm{r}}
                      e^{-i\omega t}
                      \left.\vphantom{e^{-i\omega t}} \right),
                      \label{eq:Evec_pol}
  \end{split}
\end{align}
and analogous for the $y$ component.

Eq.~\eqref{eq:Evec_pol} describes the electric field of an electromagnetic wave propagating along
the $z$ direction.
The first order beyond the dipole approximation is obtained by substituting:
\begin{align}
  e^{i\frac{\omega}{c}\hat{\bm{e}}_z\cdot\bm{r}}=1
          +i\left( \frac{\omega}{c}\hat{\bm{e}}_z\cdot\bm{r}\right)
          +\mathcal{O}\left(\left[\frac{\omega}{c}\hat{\bm{e}}_z\cdot\bm{r}\right]^2\right)
          \label{eq:multip_exp}
\end{align}
Introducing this into Eq.~\eqref{eq:Evec_pol}, we obtain:
\begin{subequations}
  \begin{align}
    E_x(\bm{r},t) & \approx E_x^{(0)}(t)+E_x^{(1)}(\bm{r},t) \tag{\theequation}\\
    &=\frac{1}{\sqrt{2\pi}}\int_{0}^{\infty} \dif \omega
                     \left(|\epsilon_x(\omega)|e^{-i\varphi_1}e^{i\omega t}
                      +|\epsilon_x(\omega)|e^{i\varphi_x}e^{-i\omega t}\right)\\
    -&\frac{i \left(\hat{\bm{e}}_z\cdot\bm{r}\right)}{c\sqrt(2\pi)}
  \int_{0}^{\infty} \dif \omega
                     \left(\omega |\epsilon_x(\omega)|e^{-i\varphi_1}e^{-i\omega t}
                      +\omega |\epsilon(\omega)|e^{-i\varphi_1}e^{i\omega t}\right)
  \end{align}
\end{subequations}
By performing a Fourier transform to time domain, we finally arrive at the
following expression,
\begin{align}
  E_x(\hat{\bm{r}},t)\approx |\epsilon_x(t)|e^{-i\varphi_x}-\frac{\left(\hat{\bm{e}}_z\cdot\hat{\bm{r}}\right)}{c}\frac{\dif |\epsilon_x(t)|}{\dif t}e^{i\varphi_x},
  \label{eq:Ex_appr}
\end{align}
which, together with one of Maxwell's equations in Fourier space,
$\bm{B}=\frac{1}{c}\bm{E}\times\hat{\bm{e}}_z$, defines
an arbitrary elctromagnetic wave in the time domain beyond the
electric dipole approximation to next highest order.

\end{document}